\def\Mearth{{\rm\,M_{Earth}}}
\def\Mjup{{\rm\,M_{Jup}}}
\def\Msat{{\rm\,M_{Sat}}}
\def\Mp{{\rm\,M_{p}}}
\def\all{{\rm\,\texttt{all}}}
\def\mixed{{\rm\,\texttt{mixed}}}
\def\mgrad{{\rm\,\texttt{mgrad}}}
\def\mequal{{\rm\,\texttt{mequal}}}
\def\mixedone{{\rm\,$\texttt{mixed1}$}}
\def\mixedtwo{{\rm\,$\texttt{mixed2}$}}
\def\jsn{{\rm\,\texttt{mgrad:JSN}}}
\def\3jjs{{\rm\,\texttt{mgrad:3JJS}}}
\def\sj3j{{\rm\,\texttt{mgrad:SJ3J}}}
\def\jup{{\rm\,\texttt{mequal:Jup}}}
\def\jthree{{\rm\,\texttt{mequal:J3}}}
\def\thirtye{{\rm\,\texttt{mequal:30e}}}
\def\nsj{{\rm\,\texttt{mgrad:NSJ}}}
\def\sat{{\rm\,\texttt{mequal:Sat}}}
\def\eg{{\it e.g.}}
\def\ie{{\it i.e.}}
\def\gsim{~\rlap{$>$}{\lower 1.0ex\hbox{$\sim$}}}
\def\lsim{~\rlap{$<$}{\lower 1.0ex\hbox{$\sim$}}}

\documentclass[12pt,preprint]{aastex}
\usepackage{graphicx}
\usepackage{longtable}
\usepackage{appendix}
\usepackage{natbib}
\usepackage{amsmath,amsthm}
\tabletypesize{\small}

\begin{document}

\title{Secular Behavior of Exoplanets: Self-Consistency and Comparisons with the Planet-Planet Scattering Hypothesis}
\author{Miles Timpe\altaffilmark{1,2,3}, Rory Barnes\altaffilmark{1,2}, Ravikumar Kopparapu\altaffilmark{4,2}, Sean N. Raymond\altaffilmark{5,2}, Richard Greenberg\altaffilmark{6}, Noel Gorelick\altaffilmark{7}}

\altaffiltext{1}{Astronomy Department, University of Washington, Box 351580, Seattle, WA 98195}
\altaffiltext{2}{Virtual Planetary Laboratory, USA}
\altaffiltext{3}{E-mail: apskier@astro.washington.edu}
\altaffiltext{4}{Pennsylvania State University, University Park, PA 16802}
\altaffiltext{5}{Laboratoire d'Astrophysique de Bordeaux, 2 Rue de l'Observatoire, BP 89 33270 Floirac, France}
\altaffiltext{6}{Lunar and Planetary Laboratory, University of Arizona, Tucson, AZ 85721}
\altaffiltext{7}{Google, Inc., 1600 Amphitheater Parkway, Mountain View, CA 94043}


\begin{abstract}
If mutual gravitational scattering among exoplanets occurs, then it may produce unique orbital properties. For example, two-planet systems that lie near the boundary between circulation and libration of their periapses could result if planet-planet scattering ejected a former third planet quickly, leaving one planet on an eccentric orbit and the other on a circular orbit. We first improve upon previous work that examined the apsidal behavior of known multiplanet systems by doubling the sample size and including observational uncertainties. This analysis recovers previous results that demonstrated that many systems lay on the apsidal boundary between libration and circulation. We then performed over 12,000 three-dimensional \emph{N}-body simulations of hypothetical three-body systems that are unstable, but stabilize to two-body systems after an ejection. Using these synthetic two-planet systems, we test the planet-planet scattering hypothesis by comparing their apsidal behavior, over a range of viewing angles, to that of the observed systems and find that they are statistically consistent regardless of the multiplicity of the observed systems. Finally, we combine our results with previous studies to show that, from the sampled cases, the most likely planetary mass function prior to planet-planet scattering follows a power law with index -1.1. We find that this pre-scattering mass function predicts a mutual inclination frequency distribution that follows an exponential function with an index between -0.06 and -0.1.
\end{abstract}

\clearpage


\section{Introduction \label{sec:intro}}

As the number of known exoplanets increases, our impressions of what constitutes a typical planetary system evolve. The discovery of each new planet and its characteristics provides new clues to the formation mechanisms and behavior of planetary systems in general. Consideration of known extrasolar systems as a population has led to new understanding of planetary evolution, from the formation of planetary systems to the death of the planets themselves.

Any formation scenario must adequately reproduce the configurations and distributions of observable systems. A number of plausible formation scenarios exist, but uncertainties remain. For example, while gas-driven migration \citep[\eg][]{LinPap86,Lin1996} likely played a role, alone it does not explain high eccentricity and/or high mutual inclination systems. Such systems may also have resulted from planet-planet scattering, in which one or more planets are ejected from a system due to mutual gravitational interactions between the planets \citep{RasioFord96,wm96}. This model has successfully produced high inclinations \citep{mw02,chatterjee08,Barnes11_upsand}, large eccentricities \citep{RasioFord96,wm96,LinIda97,Ford01,mw02,Ford08}, packed planetary systems \citep{Raymond09} and mean motion resonances \citep{Raymond08a}. Additionally, \citet{Raymond09pd1,Raymond10} showed that planet-planet scattering in the presence of planetesimal disks can reproduce planetary eccentricities at large orbital radii and the observed mass dependence of the eccentricity distribution. However, \citet{bg07a} found that the ejection of a planet was unlikely to produce the observed $\upsilon$ And system as proposed in \citet{Ford05}.

Here we reconsider the apsidal behavior of planetary systems that result from planet-planet scattering. In a coplanar non-resonant system, apsidal behavior is determined by the oscillations of its planetary eccentricities $e$ and accompanying variations of longitudes of pericenter $\varpi$, which are dynamically coupled. A significant fraction of adjacent planet pairs reside near an ``apsidal boundary,'' defined here as the boundary in orbital parameter space between apsidal libration (the difference in the longitudes of pericenter $\Delta \varpi$ oscillates about a fixed value) and circulation ($\Delta \varpi$ circulates through 360$^{\circ}$) \citep{bg06c}. In such cases, the orbital eccentricity of one planet periodically drops to near-zero. We can quantify how close the apsidal behavior is to the libration-circulation boundary using the $\epsilon$ parameter defined by \citet{bg06a},

\begin{displaymath}
\epsilon \equiv \frac{2[min(\sqrt{x^{2}+y^{2}})]}{(x_{max}-x_{min})+(y_{max}-y_{min})},
\end{displaymath}

\noindent where x and y are the Cartesian coordinates in the polar plot: $x \equiv e_{1}e_{2} \sin{(\Delta\varpi)}$; $y \equiv e_{1}e_{2} \cos{(\Delta\varpi)}$. If $\epsilon$ is less than a critical value $\epsilon_{crit}$, then the planet pair is said to be near an apsidal boundary. Here, $\epsilon_{crit}$ is defined as 0.01.

The $\epsilon$ parameter was originally conceived to describe apsidal behavior in two-planet systems dominated by secular interactions, but it is possible to calculate $\epsilon$ in systems with more than two planets, as well as those dominated by non-secular interactions. For systems with more than two planets, we calculate $\epsilon$ for each adjacent planet pair in the system. In these systems, the secular interaction between the planets results in a superposition of oscillations, with different amplitudes and frequencies \citep{murraydermott99}. If a pair resides in a system with more than two planets and $\epsilon<\epsilon_{crit}$, then the closest type of apsidal boundary must be determined by examining its motion in x-y space. 

In non-resonant two-planet systems, $\epsilon$ is expected to be small if planet-planet scattering events occur abruptly \citep{bg07a}. If the time from initial instability to the ejection event spans a small number of scattering interactions, it is likely that one of the two remaining planets will be left on a near-circular orbit. In the subsequent evolution of the system, the initially near-circular planet's eccentricity will periodically return to near-zero, and $\epsilon$ will be small. 

Early observations of the $\upsilon$ And system suggested that it was an example of near-boundary behavior, which was then shown to be a possible outcome of planet-planet scattering \citep{Malhotra02,Ford05}. However, assuming coplanar and nearly circular initial orbits, \citet{bg07a} found that planet-planet scattering only reproduced near-boundary apsidal behavior in the case of $\upsilon$ And in about 5\% of simulations.

Now, with many more known systems, we reconsider whether planet-planet scattering can reproduce the observed $\epsilon$ distribution. We compare the $\epsilon$ distribution of the observed multiplanet systems to thousands of synthetic planetary systems that were systematically generated by scattering simulations. These synthetic systems were first examined in \citet{Raymond08a} to show that planet-planet scattering can create mean motion resonances, and then in \citet{Raymond09} to demonstrate that scattering can produce packed planetary systems. 

In this study we assume that prior to scattering, the architectures of planetary systems subscribe to a few universal trends, such as planetary mass distribution and interplanetary spacing. The scattering process is stochastic, so in any individual system the post-scattering architecture cannot be easily predicted. However, certain aspects of the scattering process, such a preferential removal of small-mass planets, suggest scattering should imprint specific features in the orbital architectures of observed systems. With this picture in mind, we analyzed a large suite of \emph{N}-body integrations of initially unstable planetary systems in order to examine the orbital dynamics of stabilized two-planet systems, specifically the $\epsilon$ distribution. As we see below, certain pre-scattering architectures produce a population of stable planetary systems with properties consistent with observed multiplanet systems.

Our study consists of two parts. First we re-examine the apsidal behavior of multiplanet systems, improving on the study in \citet{bg06a} by also considering observational uncertainties. From these calculations, we build a distribution of $\epsilon$ that can be compared to theoretical predictions. Second, we analyze a large set of numerical simulations of planet-planet scattering in order to determine the $\epsilon$ distribution predicted by different initial conditions. We further introduce a mass-inclination degeneracy into our simulated data by calculating minimum masses that would be observed by radial velocity observations at different viewing angles. We can then directly compare the observed $\epsilon$ distribution to simulated $\epsilon$ distributions to constrain the properties of exoplanet systems prior to scattering. In the next section we describe our numerical and statistical methods. Finally, we predict the observed mutual inclination distribution predicted by scattering from our favored pre-scattering mass distributions. In \S\ref{sec:methods} we describe our methods. In \S\ref{sec:results} we present our results. In \S\ref{sec:discussion} we discuss the results, focusing on a joint analysis with the results of \citet{Raymond08a,Raymond09}. Finally, in \S\ref{sec:conclusions} we draw our conclusions.


\section{Methods \label{sec:methods}}


\subsection{The Observed Population \label{sec:met_observed}}

Using the best-fit orbital parameters for each observed system, and using the minimum planetary masses for the true masses, we simulate their subsequent dynamical motion and characterize their apsidal behavior. We use the symplectic \emph{N}-body integrator HNBody \citep{hnbody} to integrate the systems. The observed systems were integrated for $10^{5}$ years; nine systems that failed to complete at least one secular cycle were further integrated to $10^{6}$ years to determine the type of apsidal motion. 

The observed systems in our sample are stable on long timescales. We do not include any system for which the eccentricity of one or more planets was set to zero by observers, because these planets actually tend to have poorly constrained eccentricities \citep{shenturner2008}. The sample of observed multiplanet systems (two or more planets) includes 41 stable systems, comprising 57 adjacent planet pairs.

To quantify the effect that observational uncertainties have on our statistical comparisons, we simulate the observed systems with observational uncertainties and compare them to their corresponding best-fit cases. After simulating the published best-fit orbital parameters, we then simulate the observed systems again, but allow their initial orbital parameters to vary within their published observational uncertainties. Because the computational resources required for these simulations are immense, we do not assume any correlation between the uncertainties in different orbital parameters. For the observed simulations including uncertainties, we use a Monte Carlo approach to simulate the possible range of initial configurations for each observed system. We simulate 100 such configurations for each observed system. The initial orbital parameters in each of these configurations are selected from within each parameter's uncertainty range, assuming a normal distribution about the published best-fit value. As many observed planets lie near instability \citep{BarnesQuinn2004, bg06b, bg07b, FangMargot2012}, some of these configurations are unstable, and were thrown out. 

These 100 unique configurations for each observed system allow us to determine the range of permitted $\epsilon$ values for each planetary pair. By repeatedly cycling through the observed systems and choosing a configuration (\ie, an $\epsilon$ value) at random, we are able to create an arbitrary number of observed $\epsilon$ distributions. These distributions represent the range of possible distributions in the observed population. We generate 10,000 of these $\epsilon$ distributions. 


\subsection{Pre-Scattering Populations \label{sec:initial}}

We consider a diverse set of synthetic two-planet systems that result from unstable three-planet systems. We use the simulations of \citet{Raymond08a}, which consisted of ten sets, each with a particular mass distribution. We therefore only briefly describe them here. In the two largest sets (called \mixedone~and \mixedtwo~with 1000 simulations each), the planet masses were randomly selected from the probability distribution $\emph{dN/dM} \propto M^{-1.1}$, similar to the observed distribution \citep{butler06}. In the \mixedone~set, the planet mass $\Mp$ was restricted to fall between $\Msat$ and 3$\Mjup$. In the \mixedtwo~set, the minimum planet mass was decreased to 10$\Mearth$. Four sets (500 simulations each) were also performed, wherein all three planets were of equal mass: $\Mp$~=~30$\Mearth$, $\Msat$, $\Mjup$, and 3$\Mjup$. Finally, the $\mgrad$ sets (250 simulations each) contained radial gradients in $\Mp$ as follows: for the $\jsn$ set, in order of increasing orbital distance,$\Mp$~=~$\Mjup$, $\Msat$, and 30$\Mearth$. For the $\nsj$ set, these masses were reversed, i.e., the $\Mjup$ planet became the outermost planet. The $\3jjs$ and $\sj3j$ sets had, in increasing radial distance, $\Mp$~=~3$\Mjup$, $\Mjup$, and $\Msat$, and $\Mp$~=~$\Msat$, $\Mjup$, and 3$\Mjup$, respectively. The outermost planet in each simulation was placed two Hill radii interior to 10 AU. The simulations were designed to be initially unstable so that planet-planet scattering would occur. These systems were integrated for $10^{8}$ years with a 20 day time step using the hybrid version of the symplectic \emph{N}-body integrator MERCURY \citep{Chambers99}.

Not all of the synthetic systems above were included in our statistical analysis---as we are interested in the secular behavior, we selected only the simulations from \citet{Raymond08a} that resulted in two planets on stable orbits after scattering, leaving 1,287 synthetic systems of the original scattering simulations. 


\subsection{Mass-Inclination Degeneracy \label{sec:msini}}

For our synthetic systems, we eliminate the biasing effect of the observational mass-inclination degeneracy in a statistical sense by calculating the radial velocities as a function of viewing angle. Because radial velocity studies cannot measure mutual inclinations, for simplicity theoretical studies have usually assumed coplanar orbits. For each of the 1,287 synthetic systems, we simulate 10 different viewing geometries, each with an equal probability of being observed, which increases the number of synthetic systems in our sample to 12,870. For each viewing geometry, we scale the mass of each planet by sin\emph{i} and place the two planets' orbits in the same plane, where $90^{\circ}-i$ is the angle between the observer-star direction and the orbital plane. In essence, we impose the mass-inclination degeneracy on our synthetic sample. This approach allows for a meaningful comparison between the planet-planet scattering model (\ie, our synthetic systems) and the observed systems.

Due to the mass-inclination degeneracy, the actual value of $\epsilon$ is not known for observed systems. Therefore, we impose it on our simulated system so that our values of $\epsilon$ are directly comparable to the observed systems. Once the masses and inclinations of exoplanets are known, a revised version of $\epsilon$ may be desirable, but until then $\epsilon$ is a useful statistic to quantify the dynamical properties of radial-velocity detected planets.

We assume that a synthetic planet is detectable only if its resultant radial velocity semi-amplitude is $> 3$ m s$^{-1}$. Once an inclination is high enough that the signal is below this threshold, we no longer calculate the resultant apsidal motion, as only one planet is detectable. This produces a total of 12,674 cases.


\subsection{Calculating Apsidal Behavior \label{sec:numerical}}

The final synthetic set of 12,674 two-planet systems was integrated for $10^{5}$ years with a 20 day time step using the hybrid version of the symplectic \emph{N}-body integrator MERCURY. We integrate each of these systems in a coplanar configuration and calculate $\epsilon$. Some systems were unstable, which is not too surprising as instabilities can arise over any timescale, and were thrown out. All simulations were required to conserve energy to $dE/E < 10^{-4}$, where $dE/E$ is the change in the total energy of the system compared to its initial value. \cite{BarnesQuinn2004} showed that this requirement is adequate to test stability. Those systems with $dE/E > 10^{-4}$ were reduced to a time step of 5 days and reintegrated. If a system failed to complete a secular cycle in $10^{5}$ years, we integrated it to $10^{6}$ years. We are left with a robust set of synthetic systems which can be compared to the observed systems, see \S\ref{sec:met_observed}. This methodology produces a total of 12,674 systems from which we can compare the apsidal behavior generated by planet-planet scattering to the observed apsidal beahvior.


\subsection{Statistical Methods \label{sec:met_statistics}}

We use the well known Kolmogorov-Smirnov (K-S) test for our statistical analysis in this paper. In our comparisons of the observed and synthetic samples we utilize the two-sample K-S statistic, which tests the null hypothesis that a given synthetic planetary set is drawn from the same distribution as the observed population. For each of these comparisons, we calculate the $p$-value to determine if we can reject that hypothesis. Comparisons that result in a $p$-values below 0.01 indicates that a synthetic set does not reproduce the distribution of apsidal behavior in the observed population. 

The two-sample K-S statistic is a measure of the maximum vertical deviation between two empirical cumulative distribution functions

\begin{displaymath}
D_{n_{1},n_{2}} = Max\mid F_{n_{1}}(x) - F_{n_{2}}(x) \mid,
\end{displaymath}

\noindent where $F_{n_{1}}(x)$ and $F_{n_{2}}(x)$ are the respective cumulative distribution functions of samples $n_{1}$ and $n_{2}$. The two-sample K-S test is one of the most useful and general nonparametric methods for comparing empirical samples, as it is sensitive to differences in both location and shape of the cumulative distribution functions of the two samples.


\section{Results \label{sec:results}}


\subsection{Observed Apsidal Behavior \label{sec:res_observed}}

We begin by describing the apsidal properties of observed multiplanet systems. Table~\ref{tbl:table1} lists the values of $\epsilon$ for the current catalog of observed two-planet systems, as well as their modes of apsidal behavior: C for circulation; L$_{0}$ for aligned libration; L$_{180}$ for anti-aligned libration; C/L for small-$\epsilon$ cases, which are near the libration/circulation boundary; and C\&L for cases that alternate over time between both modes. Class identifies pairs that have undergone tidal evolution (T), are experiencing resonant interactions (R$_{n:m}$ where the $n:m$ subscript indicates the frequency ratio), or are dominated by secular interactions (S). The eccentricity evolution for each system can be found on the Extrasolar Planet Interactions website located at \texttt{http://xsp.astro.washington.edu}. In Table~\ref{tbl:table2} we list the same parameters as before in Table~\ref{tbl:table1}, but for the observed multiplanet systems with more than two planets.

Also shown in Table~\ref{tbl:table1} and Table~\ref{tbl:table2} are the mean $\epsilon$ values ($\overline{\epsilon}$) and their associated standard deviations $\sigma_{\epsilon}$. These values constrain the range of $\epsilon$ for each system within its observational uncertainty ranges and provide a self-consistency check on the observed set of systems. We perform this check because our conclusions are based on the observed systems' best-fit orbital solutions. To determine if the observed best-fit simulations are consistent with the observed simulations subject to observational uncertainties, we make a statistical comparison between each of the 10,000 $\epsilon$ distributions generated using uncertainties (see \S\ref{sec:met_observed}) and the best-fit $\epsilon$ distribution using two-sample K-S tests. Each of the 10,000 statistical comparisons results in a  $p$-value from the K-S test, and from these tests we obtain a mean $p$-value of $\overline{p} = 0.4146$ with a standard deviation of $\sigma_{\epsilon} = 0.2562$. The high mean of the $p$-value indicates that the observational uncertainties do not dramatically affect the distribution of apsidal behavior in the observed sample--- indeed, 99.9\% of the 10,000 $\epsilon$ distributions fall within 3-sigma ($p > 0.01$) of the best-fit distribution.


\begin{deluxetable}{cccccccc}
\tablecolumns{8}
\tablewidth{0pc}
\tablecaption{Apsidal Motion of Observed 2 Planet Systems \label{tbl:table1}}
\tablehead{
\colhead{Identifier} & \colhead{Pair} & \colhead{Mode} & \colhead{$\epsilon$} & \colhead{$\overline{\epsilon}$} & \colhead{$\sigma_{\epsilon}$} & \colhead{Class}
}
\startdata
BD-08$^{\circ}$2823$^{1}$ & b-c & C & 0.474 & 0.472 & 0.014 & T\\
HAT-P-13$^{2}$ & b-c & L$_{0}$ & 0.644 & 0.513 & 0.507 & S\\
HAT-P-17$^{3}$ & b-c & C & 0.412 & 0.432 & 0.030 & S\\
HD 11506$^{4}$ & c-b & C & 0.345 & 0.134 & 0.320 & S\\
HD 11964$^{5}$ & c-b & C & 0.473 & 0.474 & 0.006 & T\\
HD 12661$^{6}$ & b-c & L$_{180}$ & 0.048 & 0.083 & 0.058 & S\\
HD 17156$^{7}$ & b-c & C & 0.097 & 0.002 & 0.012 & R$_{5:1}$\\
HD 45364$^{8}$ & b-c & L$_{180}$ & 0.818 & 0.989 & 0.349 & R$_{3:2}$\\
HD 47186$^{9}$ & b-c & C & 0.494 & 0.495 & 8.34$\times$10$^{-4}$ & S\\
HD 60532$^{10}$ & b-c & L$_{180}$ & 0.078 & 0.089 & 0.060 & R$_{3:1}$\\
HD 82943$^{11}$ & b-c & C/L$_{0}$ & 1.24$\times$10$^{-3}$ & 3.49$\times$10$^{-4}$ & 1.46$\times$10$^{-10}$ & R$_{2:1}$\\
HD 99492$^{12}$ & b-c & C & 0.481 & 0.480 & 0.002 & S\\
HD 108874$^{6}$ & b-c & L$_{180}$ & 0.598 & 0.384 & 0.388 & R$_{4:1}$\\
HD 113538$^{12}$ & b-c & L$_{180}$ & 1.274 & 0.627 & 1.135 & S\\
HD 128311$^{11}$ & b-c & C/L$_{0}$ & 6.11$\times$10$^{-4}$  & 0.333 & 1.778 & R$_{2:1}$\\
HD 134987$^{13}$ & b-c & L$_{180}$ & 0.788 & 0.411 & 0.286 & T\\
HD 147018$^{14}$ & b-c & C & 0.356 & 0.349 & 0.018 & S\\
HD 155358$^{15}$ & b-c & L$_{180}$ & 0.208 & 0.180 & 0.384 & S\\
HD 163607$^{16}$ & b-c & C/L$_{180}$ & 1.01$\times$10$^{-4}$ & 0.204 & 0.143 & S\\
HD 168443$^{6}$ & b-c & C & 0.222 & 0.220 & 0.002 & S\\
HD 169830$^{11}$ & b-c & C & 0.311 & 0.318 & 0.021 & S\\
HD 177830$^{12}$ & c-b & C & 0.156 & 0.176 & 0.048 & S\\
HD 183263$^{6}$ & b-c & C/L$_{180}$ & 0.027 & 0.023 & 0.058 & S\\
HD 187123$^{6}$ & b-c & C & 0.446 & 0.449 & 0.005 & T\\
HD 190360$^{11}$ & c-b & C & 0.376 & 0.419 & 0.054 & T\\
HD 202206$^{17}$ & b-c & C & 6.99$\times$10$^{-5}$ & 4.61$\times$10$^{-5}$ & 1.51$\times$10$^{-4}$ & R$_{5:1}$\\
HD 208487$^{18}$ & b-c & C & 0.183 & 0.299 & 0.898 & R$_{7:1}$\\
HD 217107$^{6}$ & b-c & C & 0.458 & 0.458 & 0.003 & T\\
HIP 57274$^{19}$ & b-c & C & 0.459 & 0.404 & 0.078 & T\\
\enddata

\tablerefs{(1) \citealt{hebrard09}; (2) \citealt{bakos09}; (3) \citealt{howard10}; (4) \citealt{tk09}; (5) \citealt{baines09}; (6) \citealt{wright09}; (7) \citealt{short08}; (8) \citealt{correia09}; (9) \citealt{bouchy09}; (10) \citealt{desort08}; (11) \citealt{butler06}; (12) \citealt{meschiari10}; (12) \citealt{moutou10}; (13) \citealt{jones10}; (14) \citealt{segransan09}; (15) \citealt{cochran07}; (16) \citealt{giguere11}; (17) \citealt{correia05}; (18) \citealt{gregory07}; (19) \citealt{fischer11}.}

\end{deluxetable}



\begin{deluxetable}{cccccccc}
\tablecolumns{8}
\tablewidth{0pc}
\tablecaption{Apsidal Motion of Observed $>$2 Planet Systems \label{tbl:table2}}
\tablehead{
\colhead{Identifier} & \colhead{Pair} & \colhead{Mode} & \colhead{$\epsilon$} & \colhead{$\overline{\epsilon}$} & \colhead{$\sigma_{\epsilon}$} & \colhead{Class}
}
\startdata
SS$^{1}$ & J-S & C & 0.214 & 0.214 & 0 & S\\
& S-U & C & 0.022 & 0.022 & 0 & S\\
& U-N & C & 0.004 & 0.004 & 0 & S\\
47 UMa$^{2}$ & b-c & C/L$_{0}$ & 0.139 & 0.062 & 0.133 & S\\
& c-d & C/L$_{0}$ & 0.063 & 0.045 & 0.070 & S\\
55 Cnc$^{3}$ & e-b & C & 7.42$\times$10$^{-4}$ & 0.005 & 0.014 & T\\
& b-c & C & 0.053 & 0.017 & 0.042 & S\\
& c-f & C/L$_{180}$ & 0.157 & 0.157 & 0.107 & S\\
& f-d & C/L$_{0}$ & 0.175 & 0.230 & 0.065 & S\\
61 Vir$^{4}$ & b-c & C & 0.062 & 0.045 & 0.055 & T\\
& c-d & L$_{0}$ & 0.545 & 0.372 & 0.460 & S\\
HD 37124$^{5}$ & b-d & C & 0.022 & 0.007 & 0.012 & S\\
& d-c & L$_{0}$ & 0.148 & 0.029 & 0.054 & S\\
HD 38529$^{6}$ & b-d & C & 0.243 & 0.140 & 0.111 & S\\
& d-c & C & 0.288 & 0.195 & 0.132 & S\\
HD 69830$^{7}$ & b-c & C & 0.094 & 0.110 & 0.191 & T\\
& c-d & C & 0.040 & 0.083 & 0.083 & S\\
HD 74156$^{8}$ & b-d & C\&L$_{0}$ & 0.002 & 0.002 & 0.007 & S\\
& d-c & C\&L$_{0}$ & 0.002 & 0.003 & 0.012 & S\\
HD 125612$^{9}$ & c-b & C & 0.183 & 0.072 & 0.039 & T\\
& b-d & L$_{180}$ & 0.019 & 0.009 & 0.039 & R$_{6:1}$\\
HIP 14810$^{6}$ & b-c & L$_{180}$ & 0.379 & 0.262 & 0.288 & T\\
& c-d & C & 0.141 & 0.139 & 0.031 & S\\
$\upsilon$~And$^{6}$ & b-c & C & 0.020 & 0.029 & 0.011 & T\\
& c-d & L$_{0}$ & 0.114 & 0.129 & 0.028 & S\\
$\mu$~Ara$^{10}$ & c-d & C/L$_{180}$ & 5.84$\times$10$^{-4}$ & 0.026 & 0.035 & T\\
& d-b & C/L$_{0}$ & 9.71$\times$10$^{-4}$ & 0.043 & 0.059 & R$_{2:1}$\\
& b-e & C & 0.170 & 0.177 & 0.088 & S\\
\enddata

\tablerefs{(1) JPL (http://ssd.jpl.nasa.gov/txt/p\_elem\_t1.txt); (2) \citealt{gf09}; (3) \citealt{fischer08}; (4) \citealt{vogt09}; (5) \citealt{butler06}; (6) \citealt{wright09}; (7) \citealt{lovis06}; (8) \citealt{barnes08}; (9) \citealt{curto10}; (10) \citealt{pepe07}.}

\end{deluxetable}


In our sample of observed multiplanet systems, we find that a significant fraction of planetary pairs---about 1/5--- exhibit $\epsilon < 0.01$. Given that we excluded systems wherein $e=0$ for at least one planet, our results are consistent with \cite{bg06a}, who found that about 1/3 of systems are near an apsidal boundary. Therefore, despite nearly tripling the number of pairs, the fraction of systems near an apsidal boundary has remained significant. 


\begin{figure}
\plotone{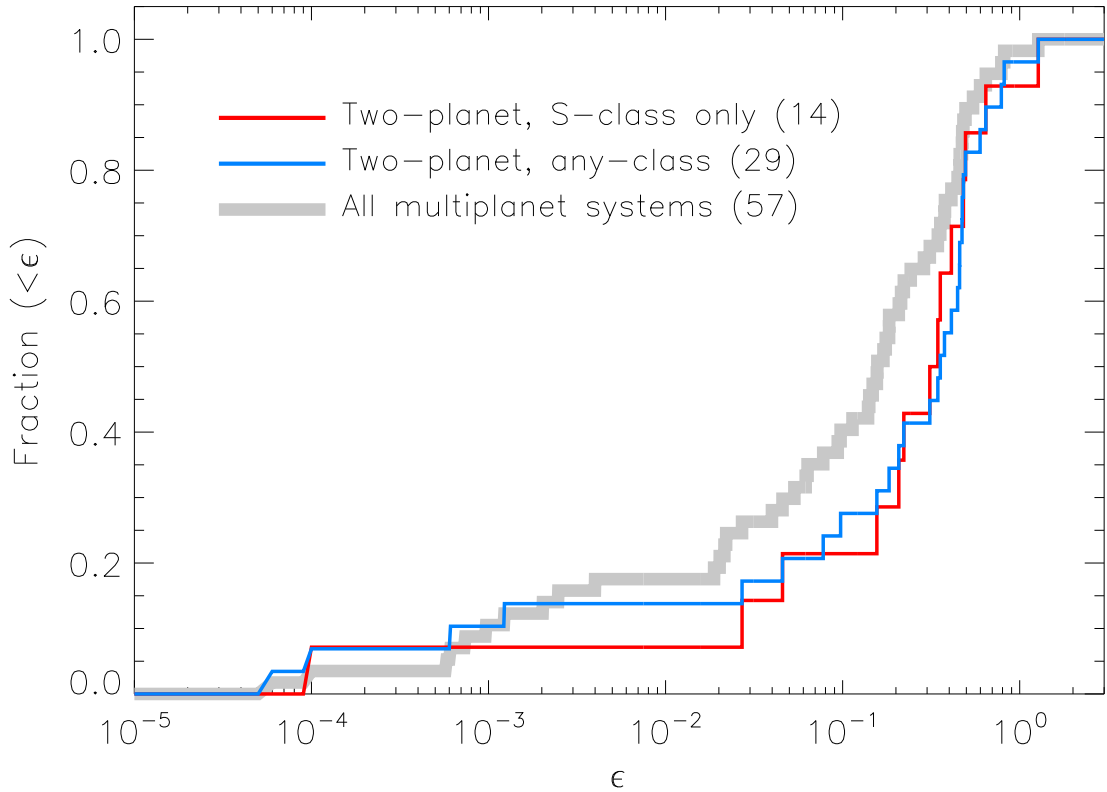}
\caption{\label{fig:one}The $\epsilon$ distribution for the different sets of the observed populations. The set of two-planet systems dominated by secular interaction is shown by the solid red line. The solid blue line also shows the $\epsilon$ distribution of two-planet systems, but without any restriction of the dominating dynamical interaction in the system. Finally, the thick gray line traces the $\epsilon$ distribution of the unrestricted set of multiplanet systems---these systems can have any number of planets and be dominated by any type of dynamical interaction (\eg, secular, resonant, or tidal).}
 \end{figure}
 

In Figure \ref{fig:one} we present the $\epsilon$ distributions for three different subsets of the observed multiplanet systems. The solid red line shows the distribution for the set that consists of strictly two-planet systems that are dynamically dominated by secular interactions. This set is important because the $\epsilon$ parameter is well-defined for such systems. We also include the $\epsilon$ distribution for the observed systems with two-planets and no restriction on dynamical interaction, as well as for the full set of all multiplanet systems. The most striking feature in Figure \ref{fig:one}, which appears in all three subsets, is a plateau in the cumulative distribution that extends from roughly $\epsilon \sim 0.0001$ to $0.02$. This plateau is indicative of a pileup of observed systems at low-$\epsilon$ values, suggesting that there is some formation mechanism or observational bias that prefers final configurations with low $\epsilon$ values. Any formation model must then reproduce not only the general distribution of $\epsilon$ values, but also the plateau in the distribution. Because all three observed distributions in Figure \ref{fig:one} are similar and exhibit the plateau feature, we consider the full set of observed planetary pairs listed in Tables~\ref{tbl:table1} and \ref{tbl:table2} for the comparisons that follow.


\subsection{Scattering-Produced Apsidal Behavior \label{sec:after}}

The synthetic systems described in \S\ref{sec:initial} provide us with an avenue to test the planet-planet scattering hypothesis against current observations. By comparing the $\epsilon$ distribution of the observed population, which we have detailed in \S\ref{sec:res_observed}, to the $\epsilon$ distributions of our scattering simulations with imposed mass-inclination degeneracy, we can determine if planet-planet scattering is capable of reproducing the observed apsidal behavior. 

\begin{figure}
\plotone{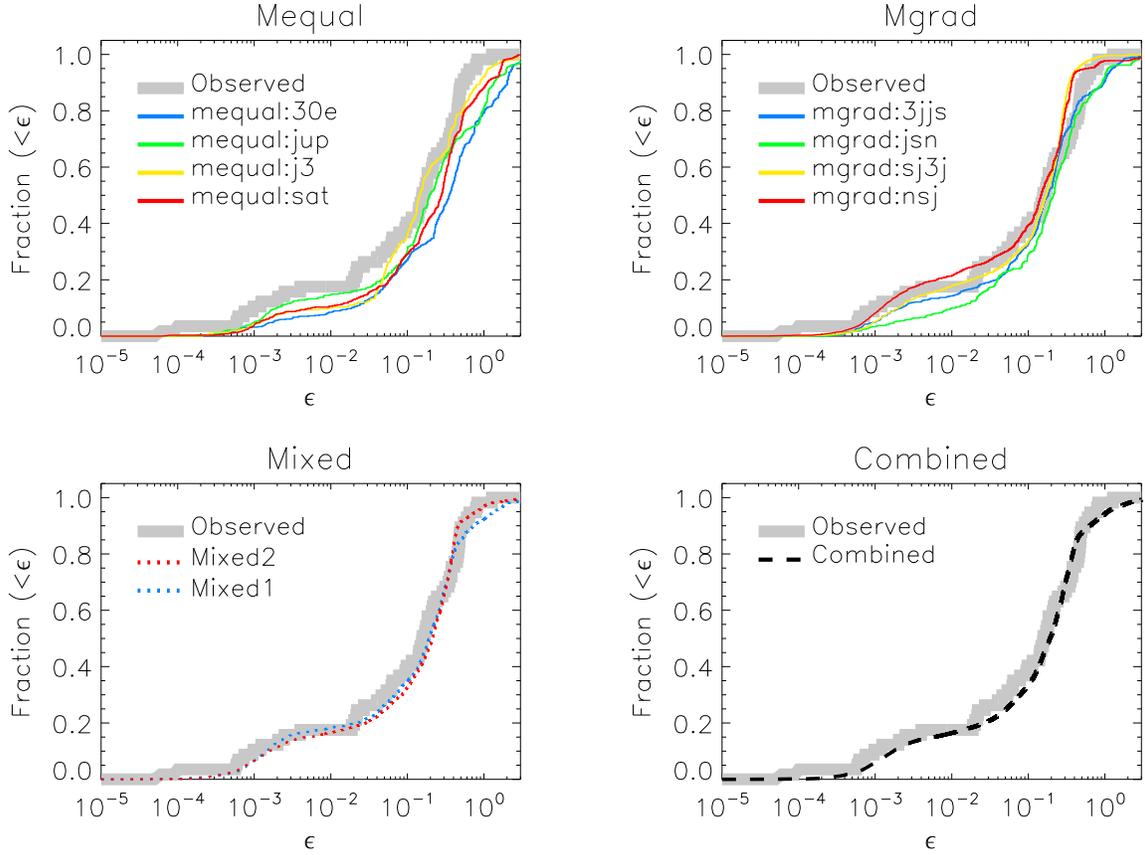}
\caption{\label{fig:two}The $\epsilon$ distribution for the best-fit orbital configurations of the observed systems (thick gray line) as compared to the $\epsilon$ distributions of the individual synthetic sets including all viewing geometries. The \mequal~sets predict to many high $\epsilon$ cases, whereas the \mgrad~sets do well reproducing the overall $\epsilon$ distribution, but don't reproduce the low-end plateau in the distribution. Of the individual synthetic sets, the \mixed~sets reproduce both the overall $\epsilon$ distribution, as well as the low-end plateau.}
 \end{figure}

Shown in Figure \ref{fig:two} are the synthetic $\epsilon$ distributions---including all viewing geometries combined into one distribution for each set---plotted against the $\epsilon$ distribution for the set of best-fit orbital configurations for the observed systems. Figure \ref{fig:two} shows the $\epsilon$ distributions for the \mequal, \mgrad, and \mixed~sets, as well as for the combined set of all synthetic systems. The combined set of synthetic systems (lower-right panel) reproduces the observed distribution very well. This similarity suggests that planet-planet scattering is capable of producing the apsidal behavior we observe in the observed population.

For all values of $\epsilon$, the individual synthetic distributions appear similar to the observed distribution, but some synthetic sets are more similar than others. The \mequal~sets appear to follow the overall observed distribution, but predict more high-$\epsilon$ systems and have trouble reproducing the low-$\epsilon$ plateau. The \mgrad~sets reproduce the low-$\epsilon$ systems, but two of the sets still fail to recreate the plateau. In contrast, both of the \mixed~sets appear to reproduce both the overall distribution and the low-$\epsilon$ plateau. Thus, upon visual inspection of Figure \ref{fig:two}, the \mixed~and \texttt{combined} sets are the most successful at reproducing the observed distribution of apsidal behavior.


\subsection{Statistical Comparison \label{sec:ks}}

We can make several different comparisons between the modeled and observed data. As we are imposing the mass-inclination degeneracy on the simulated systems, we can consider the simulated systems at all viewing geometries, or we can select one viewing geometry per system to create a simulated survey of the systems. Analogously, with the observed systems we can either consider only the best fits, or we can select at random a (stable) configuration in the permitted parameter space. In this section we consider different permutations of such samples in order to make quantitative comparisons between our synthetic planetary systems to the observed systems.

To support our visual comparisons, we provide statistical comparisons between the synthetic sets and the observed population, via two-sample Kolmogorov-Smirnov (K-S) tests. The N-body simulations of planet-planet scattering produce two-planet systems on which we impose a mass-inclination degeneracy and assume 10 different viewing geometries. We will call the synthetic configurations that result from scattering ``systems'', and the synthetic configurations that result from imposing the mass-inclination degeneracy ``simulations''. In Table~\ref{tbl:table3} and Table~\ref{tbl:table4} we present our statistical analysis, which we consider in a number of ways. In Table~\ref{tbl:table3} under "Best-Fit" we compare the synthetic simulations to a single population of observed systems (\ie, a single observed $\epsilon$ distribution), wherein the orbital parameters of each observed system are their published best-fit parameters. In Table~\ref{tbl:table4} under "Permitted" we compare the synthetic simulations to thousands of randomly generated configurations of the observed systems (\ie, thousands of observed $\epsilon$ distributions), wherein the orbital parameters of each observed system are allowed to vary within their published observational uncertainties, see \S\ref{sec:met_observed}.

In the "Best-Fit" columns we compare the best-fit configurations of the observed systems to the synthetic simulations in two ways. First we create a combined $\epsilon$ distribution for the synthetic simulations that includes all viewing geometries (\ie, a single $\epsilon$ distribution for each initial mass distribution set) and compare them to the best-fit configurations of the observed systems. The $p$-values resulting from these comparisons, $p_{BF:C}$, are shown in the leftmost column under "Best-Fit". Next we again compare the synthetic simulations to the best-fit configurations of the observed systems, but this time we choose one viewing geometry at random for each synthetic system---we repeat this method until we have 10,000 $\epsilon$ distributions for each initial mass function set that represent the full range of possible $\epsilon$ distributions resulting from different combinations of the various geometries. We perform K-S tests between each of these 10,000 synthetic $\epsilon$ distributions and the best-fit observed $\epsilon$ distribution. The right three "Best-Fit" columns show the resulting median $p$-values, $\tilde{p}_{BF:MC}$, mean $p$-values, $\overline{p}_{BF:MC}$, and the associated standard deviation, $\sigma_{\overline{p}_{_{BF:MC}}}$.

In the "Permitted" columns, the observed systems used for comparison are drawn from Monte Carlo simulations wherein the orbital parameters of each observed system are allowed to vary within their published observational uncertainties, see \S\ref{sec:met_statistics}. This gives us the statistics resulting from the permitted configurations of the observed systems. We again consider our comparison in two ways. First we randomly generate 10,000 observed $\epsilon$ distributions which represent the full range of observed $\epsilon$ distributions possible given observational uncertainties. We then compare the synthetic simulations including all geometries (\ie, a single $\epsilon$ distribution for each initial mass distribution set) to each of the 10,000 observed $\epsilon$ distributions. The resulting median $p$-values, $\tilde{p}_{P:C}$, mean $p$-values, $\overline{p}_{P:C}$, and their associated standard deviations, $\sigma_{\overline{p}_{_{P:C}}}$, are shown in the left three columns under "Permitted". Second, we compare the 10,000 synthetic $\epsilon$ distributions generated for each initial mass distribution set---via selecting one random viewing geometry for each synthetic system---to the 10,000 observed $\epsilon$ distributions resulting from the observed configurations subject to observational uncertainties. We again calculate the median $p$-value, $\tilde{p}_{P:MC}$, mean $p$-value, $\overline{p}_{P:MC}$, and the associated standard deviation, $\sigma_{\overline{p}_{_{P:MC}}}$. The right three columns under "Permitted" display these parameters. 

In the sections that follow, we quote the median $p$-values, $\tilde{p}_{BF:MC}$, obtained by comparing the best-fit configurations of the observed system to the synthetic systems with randomly chosen viewing geometries. We choose to focus on the $\tilde{p}_{BF:MC}$ statistic because the samples used in the comparison represent both a complete sampling of the synthetic $\epsilon$ distributions possible after the mass-inclination degeneracy has been imposed, and utilizes the best known orbital solutions to the observed systems. In the discussion, we quote the median $p$-values as opposed to their mean values. We thereby avoid situations in which a small fraction of simulations resulting in abnormally high $p$-values drives up the mean value.


\begin{deluxetable}{ccccccccccc}
\tablecolumns{11}
\tablewidth{0pc}
\tablecaption{Kolmogorov-Smirnov Comparisons \label{tbl:table3}}
\tablehead{
\colhead{} & \colhead{} & \colhead{} & \colhead{} & \multicolumn{5}{c}{Best-Fit} \\
\cline{5-9} \\
\colhead{} & \multicolumn{2}{c}{Set Size} & \colhead{} & \colhead{Combined} & \colhead{} & \multicolumn{3}{c}{Monte Carlo} \\
\cline{2-3} \cline{5-5} \cline{7-9} \\
\colhead{Set} & \colhead{Systems} & \colhead{Simulations} & \colhead{$\epsilon<\epsilon_{crit}$ (\%)} & $p_{_{BF:C}}$ & \colhead{} & \colhead{$\tilde{p}_{_{BF:MC}}$} & \colhead{$\overline{p}_{_{BF:MC}}$} & \colhead{$\sigma_{\overline{p}_{_{BF:MC}}}$} 
}
\startdata
\all & 1287 & 12674 & 16.5 %
& 0.6895 & & 0.5324 & 0.5276 & 0.0912 \\ %

\mixedone & 213 & 2097 & 18.6 %
& 0.6657 & & 0.8793 & 0.8341 & 0.1251 \\ %

\mixedtwo & 349 & 3448 & 16.6 %
& 0.3426 & & 0.3766 & 0.3848 & 0.0940 \\ %

\thirtye & 67 & 625 & 8.0 %
& $6 \times 10^{-4}$ & & 0.0092 & 0.0131 & 0.0091 \\ %

\sat & 66 & 630 & 11.5 %
& 0.0322 & & 0.1859 & 0.2060 & 0.1057 \\ %

\jup & 44 & 415 & 16.1 %
& 0.0079 & & 0.1981 & 0.1795 & 0.0713 \\ %

\jthree & 30 & 300 & 10.0 %
& 0.2592 & & 0.6323 & 0.6372 & 0.2070 \\ %

\nsj & 183 & 1818 & 21.5 %
& 0.0305 & & 0.0820 & 0.0829 & 0.0364 \\ %

\jsn & 53 & 530 & 10.2 %
& 0.3455 & & 0.6033 & 0.5933 & 0.1457 \\ %

\3jjs & 70 & 701 & 14.9 %
& 0.6163 & & 0.8106 & 0.7715 & 0.1377 \\ %

\sj3j & 212 & 2110 & 17.5 %
& 0.0106 & & 0.0283 & 0.0310 & 0.0150 \\ %

\textbf{Observed} & \textbf{57} & \textbf{57} & \textbf{17.5} %
& \nodata & & \nodata & \nodata & \nodata \\ %

\enddata
\end{deluxetable}



\begin{deluxetable}{ccccccccc}
\tablecolumns{9}
\tablewidth{0pc}
\tablecaption{Kolmogorov-Smirnov Comparisons \label{tbl:table4}}
\tablehead{
\colhead{} & \colhead{} & \multicolumn{7}{c}{Permitted} \\
\cline{3-9} \\
\colhead{} & \colhead{} & \multicolumn{3}{c}{Combined} & \colhead{} & \multicolumn{3}{c}{Monte Carlo} \\
\cline{3-5} \cline{7-9} \\
\colhead{Set} & \colhead{} & \colhead{$\tilde{p}_{_{P:C}}$} & \colhead{$\overline{p}_{_{P:C}}$} & \colhead{$\sigma_{\overline{p}_{_{P:C}}}$} & \colhead{} & \colhead{$\tilde{p}_{_{P:MC}}$} & \colhead{$\overline{p}_{_{P:MC}}$} & \colhead{$\sigma_{\overline{p}_{_{P:MC}}}$}
}
\startdata
\all & %
& 0.0256 & 0.0648 & 0.0983 & %
& 0.0364 & 0.0754 & 0.1002 \\ %

\mixedone & %
& 0.0309 & 0.0745 & 0.1075 & %
& 0.0669 & 0.1234 & 0.1460 \\ %

\mixedtwo & %
& 0.0298 & 0.0632 & 0.0848 & %
& 0.0491 & 0.0897 & 0.1081 \\ %

\thirtye & %
& $6 \times 10^{-5}$ & $2 \times 10^{-4}$ & $5 \times 10^{-4}$ & %
& 0.0034 & 0.0061 & 0.0080 \\ %

\sat & %
& 0.0053 & 0.0109 & 0.0167 & %
& 0.0504 & 0.0725 & 0.0705 \\ %

\jup & %
& 0.0047 & 0.0067 & 0.0073 & %
& 0.1170 & 0.1274 & 0.0758 \\ %

\jthree & %
& 0.0099 & 0.0316 & 0.0531 & %
& 0.1594 & 0.2209 & 0.1840 \\ %

\nsj & %
& 0.0349 & 0.0544 & 0.0612 & %
& 0.0735 & 0.1023 & 0.0955 \\ %

\jsn & %
& 0.0108 & 0.0312 & 0.0535 & %
& 0.1018 & 0.1413 & 0.1361 \\ %

\3jjs & %
& 0.0268 & 0.0633 & 0.0937 & %
& 0.1214 & 0.1773 & 0.1660 \\ %

\sj3j & %
& 0.0134 & 0.0249 & 0.0316 & %
& 0.0307 & 0.0476 & 0.0511 \\ %

\textbf{Observed} & %
& $\textbf{0.3139}$ & $\textbf{0.4146}$ & $\textbf{0.2562}$ & %
& $\textbf{0.3139}$ & $\textbf{0.4146}$ & $\textbf{0.2562}$ \\ %

\enddata
\end{deluxetable}


When we compared the best-fit set of observed systems to our synthetic systems with randomly chosen viewing geometries, all of our scattering sets, except \thirtye, resulted in acceptable matches---with $\tilde{p} > 0.01$. The combined suite of all ten sets resulted in a high $\tilde{p}$-value of 0.5324. Seven of the sets resulted in K-S scores of $\tilde{p} > 0.1$---\mixedone, \mixedtwo, \sat,  \jup, \jthree, \jsn, and \3jjs. The \mixedone~set produced the best match by an individual set to the observed distribution with a $\tilde{p}$-value of 0.8793.

When the uncertainties for the observed systems are included in our comparison, the \mixed~sets remain strong matches to the observed set, however the mean $p$-values for all comparisons tend to lower values. This general trend of lower statistical scores likely results from the inflation of planetary eccentricities in the catalog of observed systems--- $e$ is easily overestimated in relation to its actual value \citep{shenturner2008,zakamska2011}, whereas it is difficult to decrease $e$ without running afoul of the fundamental barrier that exists at $e=0$. Larger values of $e$ tend to inflate the $\epsilon$ value of a system, shifting the synthetic $\epsilon$ distributions away from the observed distribution and resulting in lower $\overline{p}$-values from K-S tests.

 
\section{Discussion \label{sec:discussion}}


\subsection{The $\epsilon$ Distribution \label{sec:epsilon}}

\citet{bg06c} found that the fraction of planetary pairs residing near an apsidal boundary is relatively large ($\sim1/3$), and that, in the case of coplanar and nearly circular initial orbits, planet-planet scattering is incapable of reproducing the observed configuration of the $\nu$ And system \citep{bg07a}. Our survey of known multiplanet systems confirms the findings of \citet{bg06c} that a large fraction of planetary pairs reside near an apsidal boundary (we find that $\sim 1/5$ have $\epsilon < \epsilon_{crit}$).

However, our results indicate that planet-planet scattering is capable of reproducing the apsidal behavior of observed systems. Observed systems are just as likely to exhibit dynamical motion near an apsidal boundary as our simulations of synthetic systems (17.5\% compared to 16.5\%---see Table~\ref{tbl:table3}). K-S tests show that several of our individual sets of simulations are consistent with the observed $\epsilon$ distribution. A critical factor in this agreement is that our comparisons take into account the mass-inclination degeneracy (\S\ref{sec:msini}) present in the observed set of systems. Another essential factor is the allowance of non-coplanarity in our synthetic systems---if planets are initially coplanar, scattering does not result in this agreement \citep{bg07a}.

The $\epsilon$ parameter was originally defined for two-planet systems, however we have included many observed systems with more than two planets (Table \ref{tbl:table2}) and it appears to remain valid in systems with more than two planets, as well as those undergoing all types of interactions. However, if we restrict our comparison to the observed two-planet systems, a small but significant difference arises---the fraction of systems near an apsidal boundary falls to 13.8\%, which is lower than planet-planet scattering predicts (17.5\%). The difference in the near-boundary fraction between two-planet and multiplanet systems could be a result of the relatively small statistical sample available of currently known exoplanet pairs (29 two-planet systems; 41 multiplanet systems). This possibility seems unlikely as we have confirmed the fraction of observed pairs residing near an apsidal boundary from \cite{bg06c} of 7 years ago, indicating that the distribution of observed apsidal behavior has not changed significantly since then.

On the other hand the difference could arise from the limitations of our scattering simulations. In reality, planet-planet scattering, even if the dominant formation mechanism, likely occurs in conjunction with a number of other mechanisms, such as migration or dynamical friction (\eg, interaction with gas or planetesimal disks) \citep{moeckel2008,Raymond09pd1,Raymond10,matsumura2010,moeckelarmitage2012}. Additionally, our synthetic sets are comprised entirely of two-planet systems, but it is unlikely that the average planetary system consists of only two planets. Inclusion of these processes and model systems comprised of more than two planets was beyond the scope of this study, as they require considerably more computational power, but clearly future work could examine their roles.


\subsection{Comparison With Previous Studies \label{sec:sets}}

By comparing the results of \citet{Raymond08a},  \citet{Raymond09}, and this study, which all use the same suite of scattering simulations described in \S\ref{sec:initial}, we can identify the individual sets (\ie, the pre-scattering configurations) in our simulations that most nearly reproduce the current observed configurations. A given set must perform well across all three studies to provide a compelling match to the population of observed configurations. In this section we identify and discuss those sets which perform well across all three studies.

In the first study, \citet{Raymond08a} found that four sets---\mixedone, \jup, \sat, and \thirtye---resulted in acceptable matches ($p > 0.01$) to the observed eccentricity distribution and can effectively produce mean motion resonances. In the following study, \citet{Raymond09} found that four sets---\mixedone, \mixedtwo, \sat, and \thirtye---were able to effectively create packed planetary systems. The \mixedone, \sat, and \thirtye~sets performed well in both studies, while the other sets resulted in poor matches ($p < 0.01$) to one or more characteristics of the observed population.

In \S\ref{sec:after} we showed that planet-planet scattering produces a match to the $\epsilon$ distribution, \ie, the apsidal behavior. Of the sets that performed well in \citet{Raymond08a} and \citet{Raymond09}, two of them---\mixedone and \sat---also resulted in acceptable matches ($\tilde{p} > 0.01$) in our study. While the \sat~set performed well in all three studies, a quick study of the catalog found in \citet{butler06} provides ample evidence against planetary systems comprised of equal mass planets. Thus, the \mixedone~set remains as the only synthetic set that is able to reproduce observed orbital characteristics across all three studies and put forth a realistic planetary configuration scenario. 

The \mixed~sets share the same underlying initial mass function, but are distinguished by one notable difference---the \mixedtwo~set has a significantly lower planetary mass limit (see \S\ref{sec:initial}). While \mixedone's~initial orbital configurations are robustly capable of reproducing the observed $\epsilon$ distribution, the preponderance of smaller planets in the \mixedtwo~set results in less impressive results. However, because the configurations underlying the \mixed~sets are nearly identical, we treat them together as an analog for the $\emph{dN/dM} \propto M^{-1.1}$ initial mass function and focus on the the \mixed~sets in the discussion that follows. 

These results indicate that the initial planetary mass configuration underlying the \mixed~sets produces the best match to the observed distribution of apsidal behavior. Unfortunately, the \mixed~sets (as well as all of the other scattering sets) have trouble reproducing the mass dependence of the observed eccentricity distribution. As previous studies have pointed out, high mass planets tend to have higher eccentricities than low mass planets \citep{jones2006,rme2007,fordrasio2008,wright09}. Planet-planet scattering has so far failed to reproduce this observation, despite producing a good match to the overall eccentricity distribution. \citet{Raymond10} explored the issue in depth and was able to reproduce the low and high-mass eccentricity distributions, but only through weighted combinations of \texttt{mequal} and \texttt{mgrad} sets. This approach does of course preclude the possibility of the \mixed~sets reproducing the mass dependence of the eccentricity distributions on their own. 

While we acknowledge that the \mixed~sets' failure to explain the mass dependence of the eccentricity distribution is a problem, their underlying configuration still provides the best match to a number of observed orbital characteristics across three different studies. Thus, the initial synthetic planetary configurations (\eg, the initial mass distribution, planet spacing, etc.) of the \mixed~sets provide a promising direction in which future studies could explore and hopefully refine the initial configurations of planetary systems. Indeed, the underlying initial mass distribution of the \mixed~sets ($\emph{dN/dM} \propto M^{-1.1}$) results in the best match to a number of characteristics in the the observed population, suggesting that future tests of this or similar initial planetary mass function would be wise. This result also indicates that the ordering of planets according to a specific mass gradient may not be important, as the constituent planets of the \mixed~sets are not ordered in any particular fashion.


\subsection{Predicting the Mutual Inclination Distribution \label{sec:predictions}}

\begin{figure}
\plotone{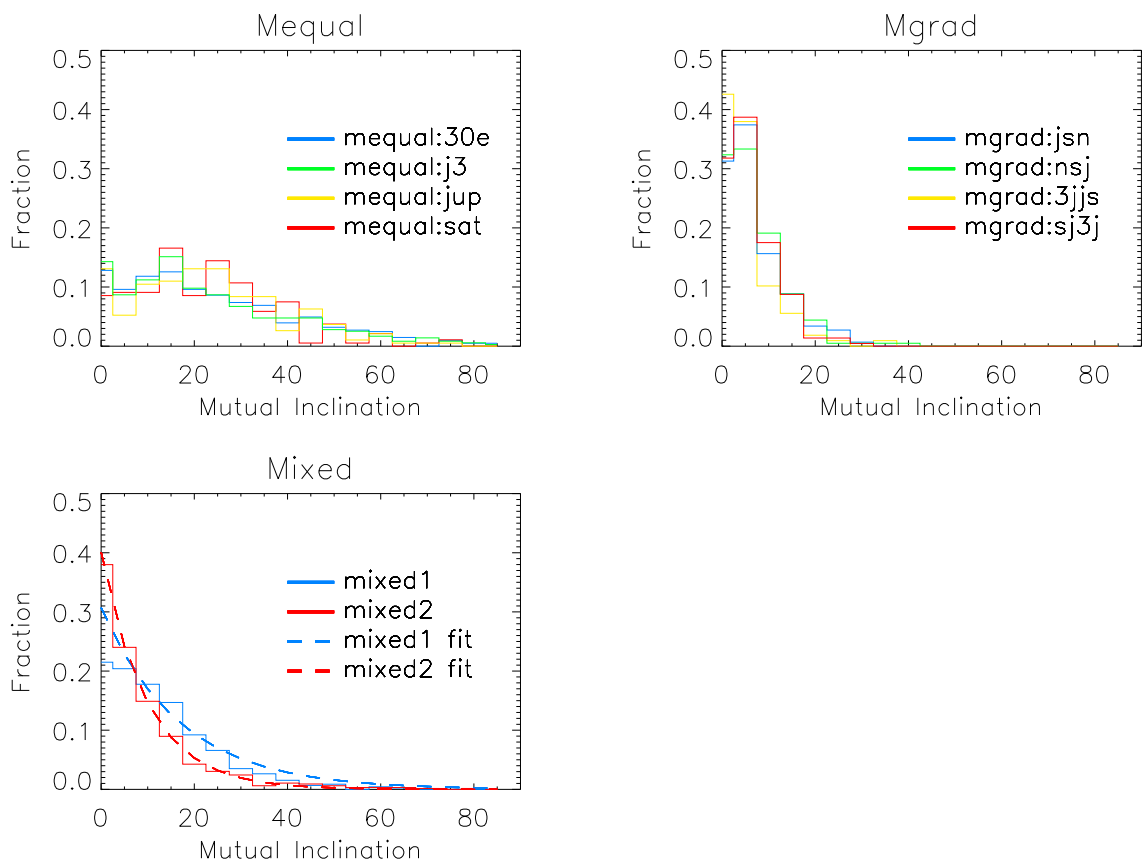}
 \caption{\label{fig:three}Distribution of mutual inclinations for our synthetic sets of systems. The \mequal~sets (\ie, systems comprised of equal mass planets) produce higher mutual inclinations than the \mgrad~sets (\ie, systems with planetary masses distributed according to a radial gradient). The fits to the \mixed~sets are exponential fits and are given in \S\ref{sec:predictions}.}
 \end{figure}

We now turn to the distribution of mutual inclinations in each of our synthetic sets. In Figure \ref{fig:three}, we provide the distributions of mutual inclinations for each of our synthetic sets. From Figure \ref{fig:three}, it is immediately apparent that the \mequal~sets result in higher typical mutual inclinations than their \mgrad~counterparts. The \mixed~sets fall comfortably in-between the \mequal~and \mgrad~sets.


Previous studies on the mutual inclination distribution of multiplanet systems have provided a wide range of results. \citet{mcarthur10} constrained the mutual inclination of $\nu$ And c and d---two planets that have been widely surveyed and studied---to be 29.9$^{\circ} \pm 1^{\circ}$ (see also \citet{reffenquirrenbach11}). The high inclinations in the $\nu$ And system appear to show that planet-planet scattering can result in high inclination orbits \citep{Barnes11_upsand}. However, drawing on the deluge of new planetary systems provided by the $Kepler$ mission, \citet{fabrycky12} found that typical mutual inclinations in the $Kepler$ ensemble lie in the range of 1.0$^{\circ}$--2.3$^{\circ}$, and concluded that planetary systems tend to be quite flat. Note, however, that $Kepler$ is unlikely to find many high mutual inclination systems through the transit method.

We predict that as observers become better equipped to constrain planetary inclinations in multiplanet systems, the observed mutual inclination distribution will shift towards higher values. Because we have highlighted the \mixed~sets as the sets with the most successful orbital configurations, we predict that the observed mutual inclination distribution will evolve towards the \mixed~mutual inclination distributions as shown in Figure \ref{fig:three}. We provide exponential fits ($F(\psi)= A e^{B \psi}$) to the \mixedone~($A=0.3071; B=-0.0594; \chi^{2}=2.519 \times 10^{-3}$) and \mixedtwo~($A=0.4010; B=-0.1012; \chi^{2}=1.642 \times 10^{-4}$) sets, where $\psi$ is the mutual inclination. The upcoming $GAIA$ mission \citep{gaia}---which will provide positional and radial velocity measurements for roughly one billion stars in the Milky Way---will be able to probe the mutual inclination distribution of extrasolar planetary systems and test the mutual inclination distribution we predict here.

The initial mass distribution in planetary systems and number of planets in a system may also play an important role in the final apsidal behavior. Our initial scattering simulations are limited to three-planet systems with a narrow range of initial planet masses. Expansion to greater than three planets, or the consideration of different mass distributions could also reproduce the observed $\epsilon$ distribution. Furthermore, the $\epsilon$ parameter's dependence on true mass provides a unique opportunity to test initial mass distributions against current observations. If simulations of systems obeying a specific initial mass distribution were to accurately recover the current $\epsilon$ distribution, then the initial mass function of planets, \ie, before the onset of scattering, could be ascertained. Such an analysis was outside the purview of this study, but should be considered in future
research.


\section{Conclusions \label{sec:conclusions}}

We have recalculated the orbital behavior of the known multiplanet systems and find that at least 1/5 of planetary pairs lie near an apsidal boundary. This result confirms the findings of \cite{bg06a}, who found that a relatively large fraction of planetary pairs lie near an apsidal boundary. Interestingly, while the $\epsilon$ parameter should only be used to characterize the apsidal behavior of planetary systems restricted to two-planets undergoing non-resonant interaction, the overall $\epsilon$ distribution does not change when it is applied to systems with more than two-planets undergoing all types of interactions. Multiplanet systems with more than two planets appear to share nearly the same distribution of apsidal behavior with two-planet systems dominated by secular interactions.

By simulating synthetic planetary systems, we were able to compare systems produced by planet-planet scattering to the observed systems. We tested these synthetic systems ability to reproduce the apsidal behavior found in observed multiplanet systems and find that planet-planet scattering is able to reproduce the distribution of apsidal behavior in the observed population. We reiterate that the set of known multiplanet systems that we use for comparison excludes any system wherein the eccentricity was set to zero by observers, thereby minimizing a source of systematic error caused by observational uncertainties. We also remove the effect of the mass-inclination degeneracy inherent in the observed population, thus allowing us to make a meaningful comparison to our synthetic scattering-produced systems.

In conjunction with previous studies using the same scattering simulations, we are able to identify the initial planetary mass distributions that most accurately reproduces the observed population. An exoplanet population with initial masses that follow a -1.1 power law produces the best fit to the observed population. While they fail to reproduce the mass dependence of the observed eccentricity distribution, their success explaining the apsidal behavior, along with similar successes explaining mean motion resonances \citep{Raymond08a} and packed planetary systems \citep{Raymond09}, make the \mixed~paradigm an interesting case with which to direct future studies.

As the best match to the observed configurations, the \mixed~sets suggest that the initial planetary mass function is described by a power law and with no identifiable ordering as function of semi-major axis. We have provided the distribution of mutual inclinations resulting from our synthetic planet-planet scattering sets, which could be used in the future as a consistency check for further studies into the pre-scattering distributions of planetary systems. Simulations such as ours continue to provide important insights into the planet formation process and explanations for observed exoplanetary systems.


\section{Acknowledgements \label{sec:acknowledgements}}
This work was funded by NSF grant AST-1108882 and by the NASA Astrobiology Institute's Virtual Planetary Laboratory lead team, under cooperative agreement No. NNH05ZDA001C. This work was facilitated through the use of advanced computational, storage, and networking infrastructure provided by the Hyak supercomputer system, supported in part by the University of Washington eScience Institute.

\bibliography{secular}

\begin{thebibliography}{65}
\expandafter\ifx\csname natexlab\endcsname\relax\def\natexlab#1{#1}\fi

\bibitem[{{Baines} {et~al.}(2009){Baines}, {McAlister}, {ten Brummelaar},
  {Sturmann}, {Sturmann}, {Turner}, \& {Ridgway}}]{baines09}
{Baines}, E.~K., {McAlister}, H.~A., {ten Brummelaar}, T.~A., {Sturmann}, J.,
  {Sturmann}, L., {Turner}, N.~H., \& {Ridgway}, S.~T. 2009, \apj, 701, 154

\bibitem[{{Bakos} {et~al.}(2009){Bakos}, {Howard}, {Noyes}, {Hartman},
  {Torres}, {Kov{\'a}cs}, {Fischer}, {Latham}, {Johnson}, {Marcy}, {Sasselov},
  {Stefanik}, {Sip{\H o}cz}, {Kov{\'a}cs}, {Esquerdo}, {P{\'a}l},
  {L{\'a}z{\'a}r}, {Papp}, \& {S{\'a}ri}}]{bakos09}
{Bakos}, G.~{\'A}., {et~al.} 2009, \apj, 707, 446

\bibitem[{{Barnes} {et~al.}(2008){Barnes}, {Go{\'z}dziewski}, \&
  {Raymond}}]{barnes08}
{Barnes}, R., {Go{\'z}dziewski}, K., \& {Raymond}, S.~N. 2008, \apjl, 680, L57

\bibitem[{{Barnes} \& {Greenberg}(2006{\natexlab{a}})}]{bg06c}
{Barnes}, R., \& {Greenberg}, R. 2006{\natexlab{a}}, \apjl, 652, L53

\bibitem[{{Barnes} \& {Greenberg}(2006{\natexlab{b}})}]{bg06a}
---. 2006{\natexlab{b}}, \apj, 638, 478

\bibitem[{{Barnes} \& {Greenberg}(2006{\natexlab{c}})}]{bg06b}
---. 2006{\natexlab{c}}, \apjl, 647, L163

\bibitem[{{Barnes} \& {Greenberg}(2007{\natexlab{a}})}]{bg07a}
---. 2007{\natexlab{a}}, \apjl, 659, L53

\bibitem[{{Barnes} \& {Greenberg}(2007{\natexlab{b}})}]{bg07b}
---. 2007{\natexlab{b}}, \apjl, 665, L67

\bibitem[{{Barnes} {et~al.}(2011){Barnes}, {Greenberg}, {Quinn}, {McArthur}, \&
  {Benedict}}]{Barnes11_upsand}
{Barnes}, R., {Greenberg}, R., {Quinn}, T.~R., {McArthur}, B.~E., \&
  {Benedict}, G.~F. 2011, \apj, 726, 71

\bibitem[{{Barnes} \& {Quinn}(2004)}]{BarnesQuinn2004}
{Barnes}, R., \& {Quinn}, T. 2004, \apj, 611, 494

\bibitem[{{Bouchy} {et~al.}(2009){Bouchy}, {Mayor}, {Lovis}, {Udry}, {Benz},
  {Bertaux}, {Delfosse}, {Mordasini}, {Pepe}, {Queloz}, \&
  {Segransan}}]{bouchy09}
{Bouchy}, F., {et~al.} 2009, \aap, 496, 527

\bibitem[{{Butler} {et~al.}(2006){Butler}, {Wright}, {Marcy}, {Fischer},
  {Vogt}, {Tinney}, {Jones}, {Carter}, {Johnson}, {McCarthy}, \&
  {Penny}}]{butler06}
{Butler}, R.~P., {et~al.} 2006, \apj, 646, 505

\bibitem[{{Chambers}(1999)}]{Chambers99}
{Chambers}, J.~E. 1999, \mnras, 304, 793

\bibitem[{{Chatterjee} {et~al.}(2008){Chatterjee}, {Ford}, {Matsumura}, \&
  {Rasio}}]{chatterjee08}
{Chatterjee}, S., {Ford}, E.~B., {Matsumura}, S., \& {Rasio}, F.~A. 2008, \apj,
  686, 580

\bibitem[{{Cochran} {et~al.}(2007){Cochran}, {Endl}, {Wittenmyer}, \&
  {Bean}}]{cochran07}
{Cochran}, W.~D., {Endl}, M., {Wittenmyer}, R.~A., \& {Bean}, J.~L. 2007, \apj,
  665, 1407

\bibitem[{{Correia} {et~al.}(2005){Correia}, {Udry}, {Mayor}, {Laskar}, {Naef},
  {Pepe}, {Queloz}, \& {Santos}}]{correia05}
{Correia}, A.~C.~M., {Udry}, S., {Mayor}, M., {Laskar}, J., {Naef}, D., {Pepe},
  F., {Queloz}, D., \& {Santos}, N.~C. 2005, \aap, 440, 751

\bibitem[{{Correia} {et~al.}(2009){Correia}, {Udry}, {Mayor}, {Benz},
  {Bertaux}, {Bouchy}, {Laskar}, {Lovis}, {Mordasini}, {Pepe}, \&
  {Queloz}}]{correia09}
{Correia}, A.~C.~M., {et~al.} 2009, \aap, 496, 521

\bibitem[{{Desort} {et~al.}(2008){Desort}, {Lagrange}, {Galland}, {Beust},
  {Udry}, {Mayor}, \& {Lo Curto}}]{desort08}
{Desort}, M., {Lagrange}, A.-M., {Galland}, F., {Beust}, H., {Udry}, S.,
  {Mayor}, M., \& {Lo Curto}, G. 2008, \aap, 491, 883

\bibitem[{{Fabrycky} {et~al.}(2012){Fabrycky}, {Lissauer}, {Ragozzine}, {Rowe},
  {Agol}, {Barclay}, {Batalha}, {Borucki}, {Ciardi}, {Ford}, {Geary}, {Holman},
  {Jenkins}, {Li}, {Morehead}, {Shporer}, {Smith}, {Steffen}, \&
  {Still}}]{fabrycky12}
{Fabrycky}, D.~C., {et~al.} 2012, ArXiv e-prints

\bibitem[{{Fang} \& {Margot}(2012)}]{FangMargot2012}
{Fang}, J., \& {Margot}, J.-L. 2012, \apj, 751, 23

\bibitem[{{Fischer} {et~al.}(2008){Fischer}, {Marcy}, {Butler}, {Vogt},
  {Laughlin}, {Henry}, {Abouav}, {Peek}, {Wright}, {Johnson}, {McCarthy}, \&
  {Isaacson}}]{fischer08}
{Fischer}, D.~A., {et~al.} 2008, \apj, 675, 790

\bibitem[{{Fischer} {et~al.}(2011){Fischer}, {Gaidos}, {Howard}, {Giguere},
  {Johnson}, {Marcy}, {Wright}, {Clubb}, {Isaacson}, {Apps}, {Lepine}, {Mann},
  {Moriarty}, {Brewer}, {Spronck}, {Schwab}, \& {Szymkowiak}}]{fischer11}
---. 2011, ArXiv e-prints

\bibitem[{{Ford} {et~al.}(2001){Ford}, {Havlickova}, \& {Rasio}}]{Ford01}
{Ford}, E.~B., {Havlickova}, M., \& {Rasio}, F.~A. 2001, \icarus, 150, 303

\bibitem[{{Ford} {et~al.}(2005){Ford}, {Lystad}, \& {Rasio}}]{Ford05}
{Ford}, E.~B., {Lystad}, V., \& {Rasio}, F.~A. 2005, \nat, 434, 873

\bibitem[{{Ford} \& {Rasio}(2008{\natexlab{a}})}]{Ford08}
{Ford}, E.~B., \& {Rasio}, F.~A. 2008{\natexlab{a}}, \apj, 686, 621

\bibitem[{{Ford} \& {Rasio}(2008{\natexlab{b}})}]{fordrasio2008}
---. 2008{\natexlab{b}}, \apj, 686, 621

\bibitem[{{Giguere} {et~al.}(2011){Giguere}, {Fischer}, {Howard}, {Johnson},
  {Henry}, {Wright}, {Marcy}, {Isaacson}, {Hou}, \& {Spronck}}]{giguere11}
{Giguere}, M.~J., {et~al.} 2011, ArXiv e-prints

\bibitem[{{Gregory}(2007)}]{gregory07}
{Gregory}, P.~C. 2007, \mnras, 374, 1321

\bibitem[{Gregory \& Fischer(2010)}]{gf09}
Gregory, P.~C., \& Fischer, D.~A. 2010, Monthly Notices of the Royal
  Astronomical Society, 403, 731

\bibitem[{{H{\'e}brard} {et~al.}(2010){H{\'e}brard}, {Udry}, {Lo Curto},
  {Robichon}, {Naef}, {Ehrenreich}, {Benz}, {Bouchy}, {Lecavelier Des Etangs},
  {Lovis}, {Mayor}, {Moutou}, {Pepe}, {Queloz}, {Santos}, \&
  {S{\'e}gransan}}]{hebrard09}
{H{\'e}brard}, G., {et~al.} 2010, \aap, 512, A46+

\bibitem[{{Howard} {et~al.}(2010){Howard}, {Bakos}, {Hartman}, {Torres},
  {Shporer}, {Mazeh}, {Kovacs}, {Latham}, {Noyes}, {Fischer}, {Johnson},
  {Marcy}, {Esquerdo}, {B{\'e}ky}, {Butler}, {Sasselov}, {Stefanik},
  {Perumpilly}, {L{\'a}z{\'a}r}, {Papp}, \& {S{\'a}ri}}]{howard10}
{Howard}, A.~W., {et~al.} 2010, ArXiv e-prints

\bibitem[{{Jones} {et~al.}(2006){Jones}, {Butler}, {Tinney}, {Marcy}, {Carter},
  {Penny}, {McCarthy}, \& {Bailey}}]{jones2006}
{Jones}, H.~R.~A., {Butler}, R.~P., {Tinney}, C.~G., {Marcy}, G.~W., {Carter},
  B.~D., {Penny}, A.~J., {McCarthy}, C., \& {Bailey}, J. 2006, \mnras, 369, 249

\bibitem[{{Jones} {et~al.}(2010){Jones}, {Butler}, {Tinney}, {O'Toole},
  {Wittenmyer}, {Henry}, {Meschiari}, {Vogt}, {Rivera}, {Laughlin}, {Carter},
  {Bailey}, \& {Jenkins}}]{jones10}
{Jones}, H.~R.~A., {et~al.} 2010, \mnras, 403, 1703

\bibitem[{{Lin} {et~al.}(1996){Lin}, {Bodenheimer}, \& {Richardson}}]{Lin1996}
{Lin}, D.~N.~C., {Bodenheimer}, P., \& {Richardson}, D.~C. 1996, \nat, 380, 606

\bibitem[{{Lin} \& {Ida}(1997)}]{LinIda97}
{Lin}, D.~N.~C., \& {Ida}, S. 1997, \apj, 477, 781

\bibitem[{{Lin} \& {Papaloizou}(1986)}]{LinPap86}
{Lin}, D.~N.~C., \& {Papaloizou}, J. 1986, \apj, 309, 846

\bibitem[{{Lo Curto} {et~al.}(2010){Lo Curto}, {Mayor}, {Benz}, {Bouchy},
  {Lovis}, {Moutou}, {Naef}, {Pepe}, {Queloz}, {Santos}, {Segransan}, \&
  {Udry}}]{curto10}
{Lo Curto}, G., {et~al.} 2010, \aap, 512, A48+

\bibitem[{{Lovis} {et~al.}(2006){Lovis}, {Mayor}, {Pepe}, {Alibert}, {Benz},
  {Bouchy}, {Correia}, {Laskar}, {Mordasini}, {Queloz}, {Santos}, {Udry},
  {Bertaux}, \& {Sivan}}]{lovis06}
{Lovis}, C., {et~al.} 2006, \nat, 441, 305

\bibitem[{{Malhotra}(2002)}]{Malhotra02}
{Malhotra}, R. 2002, \apjl, 575, L33

\bibitem[{{Marzari} \& {Weidenschilling}(2002)}]{mw02}
{Marzari}, F., \& {Weidenschilling}, S.~J. 2002, \icarus, 156, 570

\bibitem[{{Matsumura} {et~al.}(2010){Matsumura}, {Thommes}, {Chatterjee}, \&
  {Rasio}}]{matsumura2010}
{Matsumura}, S., {Thommes}, E.~W., {Chatterjee}, S., \& {Rasio}, F.~A. 2010,
  \apj, 714, 194

\bibitem[{{McArthur} {et~al.}(2010){McArthur}, {Benedict}, {Barnes},
  {Martioli}, {Korzennik}, {Nelan}, \& {Butler}}]{mcarthur10}
{McArthur}, B.~E., {Benedict}, G.~F., {Barnes}, R., {Martioli}, E.,
  {Korzennik}, S., {Nelan}, E., \& {Butler}, R.~P. 2010, \apj, 715, 1203

\bibitem[{{Meschiari} {et~al.}(2011){Meschiari}, {Laughlin}, {Vogt}, {Butler},
  {Rivera}, {Haghighipour}, \& {Jalowiczor}}]{meschiari10}
{Meschiari}, S., {Laughlin}, G., {Vogt}, S.~S., {Butler}, R.~P., {Rivera},
  E.~J., {Haghighipour}, N., \& {Jalowiczor}, P. 2011, \apj, 727, 117

\bibitem[{{Moeckel} \& {Armitage}(2012)}]{moeckelarmitage2012}
{Moeckel}, N., \& {Armitage}, P.~J. 2012, \mnras, 419, 366

\bibitem[{{Moeckel} {et~al.}(2008){Moeckel}, {Raymond}, \&
  {Armitage}}]{moeckel2008}
{Moeckel}, N., {Raymond}, S.~N., \& {Armitage}, P.~J. 2008, \apj, 688, 1361

\bibitem[{{Moutou} {et~al.}(2011){Moutou}, {Mayor}, {Lo Curto},
  {S{\'e}gransan}, {Udry}, {Bouchy}, {Benz}, {Lovis}, {Naef}, {Pepe}, {Queloz},
  {Santos}, \& {Sousa}}]{moutou10}
{Moutou}, C., {et~al.} 2011, \aap, 527, A63+

\bibitem[{{Murray} \& {Dermott}(1999)}]{murraydermott99}
{Murray}, C.~D., \& {Dermott}, S.~F. 1999, {Solar system dynamics}

\bibitem[{{Pepe} {et~al.}(2007){Pepe}, {Correia}, {Mayor}, {Tamuz}, {Couetdic},
  {Benz}, {Bertaux}, {Bouchy}, {Laskar}, {Lovis}, {Naef}, {Queloz}, {Santos},
  {Sivan}, {Sosnowska}, \& {Udry}}]{pepe07}
{Pepe}, F., {et~al.} 2007, \aap, 462, 769

\bibitem[{{Perryman} {et~al.}(2001){Perryman}, {de Boer}, {Gilmore}, {H{\o}g},
  {Lattanzi}, {Lindegren}, {Luri}, {Mignard}, {Pace}, \& {de Zeeuw}}]{gaia}
{Perryman}, M.~A.~C., {et~al.} 2001, \aap, 369, 339

\bibitem[{{Rasio} \& {Ford}(1996)}]{RasioFord96}
{Rasio}, F.~A., \& {Ford}, E.~B. 1996, Science, 274, 954

\bibitem[{{Rauch} \& {Hamilton}(2002)}]{hnbody}
{Rauch}, K.~P., \& {Hamilton}, D.~P. 2002, in Bulletin of the American
  Astronomical Society, Vol.~34, AAS/Division of Dynamical Astronomy Meeting
  \#33, 938--+

\bibitem[{{Raymond} {et~al.}(2009{\natexlab{a}}){Raymond}, {Armitage}, \&
  {Gorelick}}]{Raymond09pd1}
{Raymond}, S.~N., {Armitage}, P.~J., \& {Gorelick}, N. 2009{\natexlab{a}},
  \apjl, 699, L88

\bibitem[{{Raymond} {et~al.}(2010){Raymond}, {Armitage}, \&
  {Gorelick}}]{Raymond10}
---. 2010, \apj, 711, 772

\bibitem[{{Raymond} {et~al.}(2008){Raymond}, {Barnes}, {Armitage}, \&
  {Gorelick}}]{Raymond08a}
{Raymond}, S.~N., {Barnes}, R., {Armitage}, P.~J., \& {Gorelick}, N. 2008,
  \apjl, 687, L107

\bibitem[{{Raymond} {et~al.}(2009{\natexlab{b}}){Raymond}, {Barnes}, {Veras},
  {Armitage}, {Gorelick}, \& {Greenberg}}]{Raymond09}
{Raymond}, S.~N., {Barnes}, R., {Veras}, D., {Armitage}, P.~J., {Gorelick}, N.,
  \& {Greenberg}, R. 2009{\natexlab{b}}, \apjl, 696, L98

\bibitem[{{Reffert} \& {Quirrenbach}(2011)}]{reffenquirrenbach11}
{Reffert}, S., \& {Quirrenbach}, A. 2011, \aap, 527, A140

\bibitem[{{Ribas} \& {Miralda-Escud{\'e}}(2007)}]{rme2007}
{Ribas}, I., \& {Miralda-Escud{\'e}}, J. 2007, \aap, 464, 779

\bibitem[{{S{\'e}gransan} {et~al.}(2010){S{\'e}gransan}, {Udry}, {Mayor},
  {Naef}, {Pepe}, {Queloz}, {Santos}, {Demory}, {Figueira}, {Gillon},
  {Marmier}, {M{\'e}gevand}, {Sosnowska}, {Tamuz}, \& {Triaud}}]{segransan09}
{S{\'e}gransan}, D., {et~al.} 2010, \aap, 511, A45+

\bibitem[{{Shen} \& {Turner}(2008)}]{shenturner2008}
{Shen}, Y., \& {Turner}, E.~L. 2008, \apj, 685, 553

\bibitem[{{Short} {et~al.}(2008){Short}, {Welsh}, {Orosz}, \&
  {Windmiller}}]{short08}
{Short}, D., {Welsh}, W.~F., {Orosz}, J.~A., \& {Windmiller}, G. 2008, ArXiv
  e-prints

\bibitem[{{Tuomi} \& {Kotiranta}(2009)}]{tk09}
{Tuomi}, M., \& {Kotiranta}, S. 2009, \aap, 496, L13

\bibitem[{{Vogt} {et~al.}(2010){Vogt}, {Wittenmyer}, {Butler}, {O'Toole},
  {Henry}, {Rivera}, {Meschiari}, {Laughlin}, {Tinney}, {Jones}, {Bailey},
  {Carter}, \& {Batygin}}]{vogt09}
{Vogt}, S.~S., {et~al.} 2010, \apj, 708, 1366

\bibitem[{{Weidenschilling} \& {Marzari}(1996)}]{wm96}
{Weidenschilling}, S.~J., \& {Marzari}, F. 1996, \nat, 384, 619

\bibitem[{{Wright} {et~al.}(2009){Wright}, {Upadhyay}, {Marcy}, {Fischer},
  {Ford}, \& {Johnson}}]{wright09}
{Wright}, J.~T., {Upadhyay}, S., {Marcy}, G.~W., {Fischer}, D.~A., {Ford},
  E.~B., \& {Johnson}, J.~A. 2009, \apj, 693, 1084

\bibitem[{{Zakamska} {et~al.}(2011){Zakamska}, {Pan}, \& {Ford}}]{zakamska2011}
{Zakamska}, N.~L., {Pan}, M., \& {Ford}, E.~B. 2011, \mnras, 410, 1895

\end{thebibliography}

\appendix

\end{document}